\documentclass[amstex,aps,amsfonts,prsty]{article} 
\usepackage[dvips]{graphicx}
\usepackage{amssymb}
\usepackage{amsmath}
\usepackage{a4wide}
\usepackage{citesort}

\begin{document}

\begin{titlepage}

\begin{center}

{\LARGE \bf Ultra-Slow Vacancy-Mediated 
Tracer Diffusion in Two Dimensions: 
The Einstein Relation Verified.}

\vspace{0.3in}

{\Large \bf O.B{\'e}nichou$^{1}$ and  G.Oshanin$^2$}

\vspace{0.1in}

{\large \sl $^1$ Laboratoire de Physique de la Mati{\`e}re Condens{\'e}e, \\
Coll{\`e}ge de France, 11 Place M.Berthelot, 75252 Paris Cedex 05, France
}

{\large \sl $^2$ Laboratoire de Physique Th{\'e}orique des Liquides, \\
Universit{\'e} Paris 6, 4 Place Jussieu, 75252 Paris, France
}

\begin{abstract}
We study the dynamics of a charged tracer particle (TP)
on a two-dimensional lattice all sites of which except one (a vacancy) are 
filled with identical neutral, hard-core particles. 
The particles move randomly 
by exchanging their positions with the vacancy, subject to 
the hard-core exclusion.
In case when the charged TP experiences a bias due to external electric 
field ${\bf E}$, (which favors its jumps in the
preferential direction), 
we determine exactly
the limiting
probability distribution of the TP position 
in terms of appropriate scaling variables 
and the leading large-$n$ ($n$ being the discrete time)
behavior of the TP mean displacement $\overline{{\bf X}}_n$; the latter is shown to obey   
an
anomalous, logarithmic law  $|\overline{{\bf X}}_n| = \alpha_0(|{\bf E}|) \ln(n)$. 
On comparing our results  with earlier predictions by
 Brummelhuis and Hilhorst (J. Stat. Phys. {\bf 53}, 249
(1988)) for
 the TP diffusivity $D_n$ in the unbiased case, we infer that the Einstein relation $\mu_n = \beta
D_n$ between the TP diffusivity and the mobility $\mu_n = \lim_{|{\bf E}| \to 0}(|\overline{{\bf X}}_n|/| {\bf E} |n)$ 
holds in the leading in $n$ order, despite the fact that both  $D_n$ and  $\mu_n$ are not constant but 
vanish as $n \to \infty$. We also generalize our approach 
to the situation with very small but finite vacancy concentration $\rho$, 
in which case we
find a ballistic-type law $|\overline{{\bf X}}_n| = \pi \alpha_0(|{\bf E}|) \rho n$. 
We demonstrate that here, again, both  $D_n$ and  $\mu_n$,
calculated in the linear in $\rho$ approximation, do obey the Einstein relation.
\end{abstract}

\end{center}

\end{titlepage}


\section{Introduction.}

Consider a square lattice of which each site except one is filled with a hard-core
particle. The empty site is referred to as a "vacancy". The particles move randomly
on the lattice, their   
random walks being constrained by the condition that each site can be at most singly 
occupied. More specifically, at each moment of time $n = 1, 2, 3, \ldots$ one particle
selected with probability $1/4$ among the four particles surrounding the vacancy will
exchange its position with the vacancy.  
Suppose next that one 
selects 
one of the
particles, "tags" it and follows its trajectory ${\bf X}_n$. 
Evidently, dynamics of the tagged - the tracer particle (TP) will be quite complicated,
in contrast to the standard, by definition,  
 lattice random walk executed by the vacancy: The TP can move only when encountered by
the vacancy and its successive moves will be correlated, since the vacancy will always have a greater probability to
return for its next encounter from the direction it has left than from a perpendicular or opposite direction. 
On the other hand, it is
clear that on a two-dimensional lattice the TP will make infinitely long excursions as $n \to
\infty$ even in the presence of a single vacancy, since its random walk is
recursive in 2D and the vacancy is certain to encounter 
the tracer particle many times. 
A natural question is, of course, what are 
the statistical properties  of the TP random walk,  its mean-square displacement
$\overline{{\bf X}_n^2}$ 
from its initial position at time moment $n$, and the
probability $P_n^{(tr)}(\bf{X})$ that at time $n$ the TP
appears at position ${\bf X} = (x_1,x_2)$?

The just described model, which represents, in fact, one of the simplest cases
of the so-called "slaved
diffusion processes", has been studied  over the years in various guises, ranging from
the "constrained dynamics" model of Palmer \cite{palmer}, vacancy-mediated
bulk diffusion
in metals and crystals (see, e.g. \cite{nakazato,lebowitz2,kehr,hughes,bur2,benichou}), frictional properties of
dynamical percolative environments \cite{granek,klafter}
or dynamics of impure atoms in close-packed surfaces of metal crystals, such as, e.g., a copper 
\cite{saar1,saar2,saar3}. Brummelhuis and Hilhorst \cite{hilhorst1} were first to
present an  exact solution of this model in the
lattice formulation. It has been shown that in the presence of a
single vacancy the TP trajectories are remarkably confined;   the
mean-square displacement shows an unbounded growth, but it does 
grow only $logarithmically$ with time,
\begin{equation}
\label{log}
\overline{{\bf X}_n^2} \sim \frac{\ln(n)}{\pi (\pi - 1)}, \;\;\; \text{as $n \to\infty$},
\end{equation}     
which implies that the TP diffusivity $D_n$, defined as 
\begin{equation}
\label{Di}
D_n = \frac{\overline{{\bf X}_n^2}}{4 n} \sim \frac{\ln(n)}{4 \pi (\pi - 1) n},
\end{equation}
is not constant but rather vanishes as time $n$ progresses.
 
Moreover, it has been found \cite{hilhorst1} 
that at sufficiently large times   $P_n^{(tr)}(\bf{X})$ converges to a limiting form as a
function of the scaling variable $\eta =  |{\bf X}|/\sqrt{\ln(n)}$. Still striking,
this limiting distribution is not a $Gaussian$ but a modified Bessel
function $K_0(\eta)$, which signifies that the successive steps of the TP, although
separated by long time intervals, are effectively correlated. These results have been
subsequently reproduced by means of different analytical techniques in 
Refs.\cite{newman} and \cite{toro,zia}.

  Brummelhuis and Hilhorst have also
generalized their analytical approach to the case of a 
very small but finite vacancy concentration $\rho$
\cite{hilhorst2}, in which case a conventional diffusive-type behavior
\begin{equation}
\label{diff}
\overline{{\bf X}_n^2} = \frac{\rho \; n}{(\pi - 1)}, \;\;\; \rho \ll 1, \;\;\; n \to \infty, 
\end{equation}
has been recovered. Note that Eq.(\ref{diff}) coincides with the
earlier result of Nakazato and Kitahara \cite{nakazato} in the limit $\rho \ll 1$, and  
is well confirmed by numerical simulations \cite{kehr,ajay}.

This paper is devoted to the following, rather fundamental to our point, problem: Suppose that we charge 
the tracer particle, (while the rest are kept neutral), and switch on an electric field $\bf{E}$. 
In such a situation, the TP will have asymmetric hopping probabilities and
in its exchanges with the vacancy, depending on the TP and vacancy relative orientation,  
the TP will have a preferency (or, on contrary, a reduction of the rate) 
for exchanging its position with the vacancy compared to other three neighboring
particles. One might expect that in this case the TP mean displacement $\overline{{\bf X}}_n$  
will not be exactly equal to zero
and might define the TP mobility as
\begin{equation}
\label{mob}
\mu_n = \lim_{|\bf{E}| \to 0} \frac{|\overline{{\bf X}_n}|}{|{\bf E}| n}.
\end{equation}
Now, the question is whether the mobility $\mu_n$, calculated from the TP mean displacement in the presence of an external
electric field, and the diffusivity  $D_n$, Eq.(\ref{Di}), deduced from the TP mean-square displacement in the absence of
the field, obey the generalized Einstein relation of the form
\begin{equation}
\label{einstein}
\mu_n = \beta D_n,
\end{equation} 
where $\beta$ denotes the reciprocal temperature? 

Note that this question 
has been already addressed within the context
of the TP diffusion in one-dimensional hard-core lattice gases with arbitrary finite
vacancy concentration \cite{lebowitz2,bur2,benichou,olla,bur3}. It has been found that 
Eq.(\ref{einstein}) holds 
not only for the TP diffusion in a 1D hard-core gas on a finite lattice 
\cite{lebowitz2},
 but also for infinite 1D lattices with non-conserved \cite{benichou} and conserved particles number
\cite{bur2,olla,bur3}. Remarkably, in the latter 
 case Eq.(\ref{einstein}) holds for $n$ sufficiently large despite the fact that both the TP mobility and the
diffusivity are not constant as $n \to \infty$ but
all vanish in proportion to $1/\sqrt{n}$ \cite{bur2,olla,bur3}. 
On the other hand, it is well known that the Einstein relation is violated in some physical situations; for instance, it is
not fulfilled 
for Sinai diffusion \cite{sinai} or  
diffusion on percolation clusters, due to effects
of strong temporal trapping in the dangling ends (see also Refs.\cite{cugl1} and \cite{cugl2} for some other examples). Hence,    
in principle, 
 it is not $\it a \; priori$ clear
whether  Eq.(\ref{einstein}) should be valid for the model under study; 
here, the TP walk procceds only due to encounters 
with a $single$ vacancy, its
mean-square displacement grows only $logarithmically$ with time and
the diffusivity follows much faster decay law in Eq.(\ref{Di}), compared to 
the $D_n \sim n^{-1/2}$ law obtained for the
one-dimensional systems with finite vacancy concentrations.

The paper is structured as follows: In section 2 we present more precise formulation of the problem and introduce basic
notations.  In Section 3 we discuss our general approach to computation of the probability 
$P^{(tr)}_n(\bf{X})$ of finding the TP at position ${\bf X}$ at time moment $n$ and to evaluate 
 $P^{(tr)}_n(\bf{X})$ in the  
general form as a function of some return probabilities describing the random walk executed by the vacancy. 
The Section 4 is devoted to calculation of these return probabilities in the general case, as well as to the derivation 
of explicit expressions 
determining their asymptotical behavior. In Section 5, we present explicit asymptotical results for both the
probability distribution and the TP mean displacement. We show that as $n \to \infty$, $P^{(tr)}_n(\bf{X})$ written in terms of two
appropriate scaling variables, converges to a rather
 unusual limiting distribution. 
We also demonstrate here that the TP mobility, which is obtained
in the present work in the leading in $n$ order, and the TP diffusivity in the unbiased 
case, calculated earlier by Brummelhuis and Hilhorst
\cite{hilhorst1}, do obey the Einstein relation. Further on, in Section 6 we extend our approach to the situation with
very small but finite vacancy concentration and determine, in the leading in $n$ order, the TP mobility. We show that
also in this case the TP mobility and the TP diffusivity in the unbiased case do obey the Einstein relation, 
in the linear in $\rho$ approximation and
in the leading in $n$ order. Finally, in Section 7, we conclude with a brief summary and
discussion of our results.

\section{The model.}

Consider a two-dimensional, infinite in both  $x_1$ and $x_2$ directions, 
square lattice every site of which except one (a
vacancy)
is filled by identical hard-core particles (see Fig.1). All particles except one 
are electrically neutral. The charged particle, which is
initially at the origin,  will be referred to in what 
follows as the tracer particle - the TP. Its position at the lattice 
at time $n$  will be denoted by ${\bf X}_n$. Electric field ${\bf E}$ of strength $E = |{\bf E}|$ is
 oriented in the positive $x_1$ direction. For simplicity, the charge of the TP is set equal to unity.

Next, we suppose that at each tick of the clock, $n = 1, 2, 3, \ldots$, each particle 
selects at random a jump direction
and attempts to hop onto the target site. Evidently, the jump event 
can be only successful  for four particles adjacent to
the vacancy. 

The form of the jump direction probabilities depends on 
whether the particle is charged or not. For uncharged particles all
hopping directions are equally probable and hence, all jump direction probabilities are equal to $1/4$. On the other hand, 
the charged particle - the TP, "prefers" to jump in
the direction of the applied electric field; the normalized jump direction probabilities of 
the TP are given, in a usual fashion, by   
\begin{equation}
p_\nu= Z^{-1} \exp\Big[\frac{\beta}{2}({\bf E \cdot e}_{\boldsymbol
\nu})\Big],
\label{defp}
\end{equation} 
where $Z$ is the normalization constant, 
${\bf e}_{\boldsymbol \nu}$ is the unit vector denoting the 
jump direction,  $\nu\in\{\pm1,\pm2\}$, and  $({\bf E \cdot e_{\nu}})$ stands for the scalar product. 
We adopt the notations ${\bf
e_{\pm1}}=(\pm1,0)$ and ${\bf e_{\pm 2}}=(0,\pm1)$, which means that  ${\bf
e_{1}}$ (${\bf
e_{-1}}$) is the unit vector in the positive (negative) $x_1$-direction, while ${\bf
e_{2}}$ (${\bf e_{- 2}}$) is the unit vector in the positive (negative) $x_2$-direction. 
Consequently, the normalization constant $Z$ is
\begin{equation}
\label{zo}
Z = \sum_{\mu}\exp\Big[\frac{\beta}{2}({\bf E \cdot e}_{\boldsymbol \mu})\Big],
\end{equation}
where
the sum with the subscript $\mu$ denotes summation over all possible
orientations
of the vector ${\bf e}_{\mu}$; that is, $\mu = \{\pm1,\pm2\}$. Note that the jump direction probabilities defined by Eqs.(\ref{defp}) and
(\ref{zo}) 
do preserve the detailed balance condition.
\begin{figure}[ht]
\begin{center}
\includegraphics*[scale=0.5]{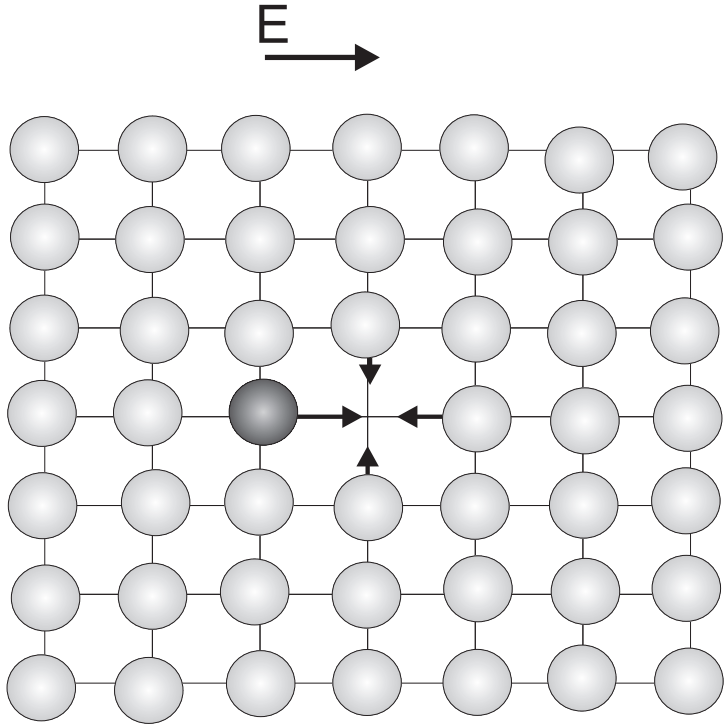}
\caption{\label{Fig3} {\small  Two-dimensional, infinite in both directions,
 square lattice in which all sites except one are filled with identical hard-core
particles (grey spheres). The black sphere denotes a single tracer particle, which is
subject to external field $\bf{E}$, oriented in the positive $x_1$ direction, and
thus has asymmetric hopping propabilities.}}
\end{center}
\end{figure}  

Next, it is expedient to  reformulate the dynamics between two consequtive 
jumps of the TP in terms of the random walk executed by the vacancy and its  
jump direction probabilities $q_{\nu}$. From the viewpoint of the vacancy, 
the jump direction probabilities depend on whether the TP is one of four surrounding particles or not. 
Evidently, in case when all four surrounding particles are electrically neutral, 
we still have that on the next time step 
the vacancy will change its position with one of four neighboring particles 
selected at random with equal probabilities. 
Hence, in case when the TP is not adjacent to the vacancy, 
all four jump directions for the vacancy are equally probable, i.e. $q_{\nu} = 1/4$.
On the other hand, the situation is a bit more complex 
when one of these four particles is the TP, which has asymmetric jump direction probabilities, Eq.(\ref{defp}). 
A natural choice of the $normalized$\footnote{Note, that normalization here insures that the vacancy performs one jump
each time step. Otherwise, we will introduce artificial "temporal trapping" probability, 
which would definitely 
lead to the violation of Eq.(\ref{einstein}).} jump direction probabilities of the vacancy 
in this case 
is as follows:      
Suppose that at time moment $n$ the tracer particle is at position ${\bf X}_n$ and the vacancy occupies an adjacent site
 ${\bf X}_n + {\bf e}_{\nu}$. Then, an exchange of the positions between the TP and 
the vacancy, which implies that the TP is moved one step in the ${\bf e_{\nu}}$-direction, takes place with the probability
\begin{equation}
\label{45}
q_{-\nu} = Z^* \; p_{\nu}
\end{equation}
while the probability of 
the exchange of positions with any of other three adjacent particles is given by
\begin{equation}
\label{46}
q_{\mu \neq -\nu} = \frac{1}{4} Z^* 
\end{equation} 
The normalization constant $Z^*$ in this case is, evidently, 
\begin{equation}
\label{zzo}
Z^* = 3/4 + p_{\nu},
\end{equation}
where $p_{\nu}$ has been defined previously in Eq.(\ref{defp}).

Consequently, apart of four sites in the immediate vicintiy of the tracer particle, the vacancy
performs a standard, symmetric random walk. 
In the vicinity of the TP, the vacancy jump direction probabilities are perturbed by
the TP asymmetric hopping rules. 
Hence, the random walk executed by the vacancy can be thought off as a particular case of the
so-called   
``random walk with defective sites'' (see Ref.\cite{hughes} for more details), 
or 
as a realization of the ``random walk with a hop-over site'' \cite{zia}.

\section{Probability distribution function $P^{ (tr)}_n({\bf X})$.}

A standard
 approach to define the properties of the TP random walk would be to start with a master 
equation determining the evolution of the whole
configuration of particles. In doing so, 
similarly to the analysis of the tracer diffusion on 2D lattices in the presence of a finite vacancy
concentration (see, e.g. Ref.\cite{benichou2}), one obtains the evolution of the joint distribution  $P_n({\bf X},{\bf Y})$ of the TP position ${\bf
X}$ and of the
vacancy position ${\bf Y}$ at time moment $n$. The property of interest, i.e. the reduced distribution function 
of 
the TP alone  will then be found from $P_n({\bf X},{\bf Y})$ by performing lattice summation over all possible values of the variable ${\bf Y}$.

Here we pursue, however, a different approach, which has 
been first put forward in the original work  of  
Brummelhuis and Hilhorst \cite{hilhorst1}; that is,  we   
construct the distribution function of the TP position at time $n$  directly
in terms of the return probabilities of the random walk performed by the vacancy. 
The only complication, compared to the unbiased case considered by
Brummelhuis and Hilhorst \cite{hilhorst1}, is that in our case  ten  different 
return probabilities would be
involved, instead of three different ones appearing  in the unbiased case. 
Hence, the analysis will be slightly more involved.

We begin by introducing some basic
notations. Let
\begin{itemize}
\item $P^{ (tr)}_n({\bf X})$ be the probability that the TP, which starts its random walk at the origin, appears 
at the site ${\bf X}$ at time moment $n$, given that the vacancy is initially at site ${\bf Y}_0$.
\item $F^*_n({\bf 0}\;|\;{\bf Y_0})$ be the probability that the
vacancy, which starts its random walk at the site ${\bf Y}_0$,
 arrives at the origin  ${\bf 0}$ for the first time at the time step $n$.
\item $F^*_n({\bf 0}\;|\;{\bf e}_{\boldsymbol \nu}\;|\;{\bf Y_0})$ be the
conditional probability that the vacancy, which starts its random walk at the site ${\bf Y}_0$, 
 appears  at the origin for the first
time at the time step $n$, being at time moment $n - 1$ at the site 
${\bf e}_{\boldsymbol \nu}$. 
\end{itemize}
Further on, for any time-dependent quantity $L_n$ we define the
generating function of the form:
\begin{eqnarray}
\label{L}
L(\xi)=\sum_{n=0}^{+\infty}L_n \xi^n
\end{eqnarray}
and for any space-dependent quantity $Y({\bf X})$ the discrete Fourier
transform
\begin{eqnarray}
\label{Y}
{\widetilde Y}({\bf k})=\sum_{\bf X} \exp\Big( i ({\bf k} \cdot {\bf X}) \Big) Y({\bf X}),
\end{eqnarray}
where the sum runs over all lattice sites.

Now, following Brummelhuis and Hilhorst \cite{hilhorst1}, we write down directly 
the equation obeyed by the reduced probability distribution 
$P^{ (tr)}_n({\bf X})$ (cf Ref.\cite{toro} for a study of the joint
probability of the TP position and of the vacancy position in the unbiased case):
\begin{eqnarray}
P^{(tr)}_n({\bf X})&=& \delta_{{\bf X},{\bf 0}}\left(1-\sum_{j=0}^n F^*_j({\bf
0}|{\bf Y_0})\right) + \nonumber\\ 
&+&\sum_{p=1}^{+\infty}\sum_{m_1=1}^{+\infty}\ldots\sum_{m_p=1}^{+\infty}
\sum_{m_{p+1}=0}^{+\infty}\delta_{m_1+\ldots+m_{p+1},n}\sum_{\nu_1}\ldots\sum_{\nu_p}\delta_{{\bf
e}_{{\boldsymbol \nu}_1}+\ldots+{\bf
e}_{{\boldsymbol \nu}_p},{\bf X}} \times \nonumber\\
&\times& \left(1-\sum_{j=0}^{m_{p+1}}F^*_j({\bf
0}|-{\bf e}_{{\boldsymbol \nu}_{\bf p}})\right) \times \nonumber\\
&\times& F^*_{m_p}({\bf 0}|{\bf e}_{{\boldsymbol \nu}_{\bf p}}\;|\;-{\bf
e}_{{\boldsymbol \nu}_{\bf p-1}})\ldots F^*_{m_2}({\bf 0}|{\bf e}_{{\boldsymbol
\nu}_{\bf 2}}\;|\;-{\bf e}_{{\boldsymbol \nu}_{\bf 1}}) F^*_{m_1}({\bf 0}|{\bf
e}_{{\boldsymbol \nu}_{\bf 1}}\;|\;{\bf Y_0}).
\label{Ptr}
\end{eqnarray}
Next, using the definition of the generating functions and of the discrete Fourier transforms, Eqs.(\ref{L}) and (\ref{Y}), 
we obtain 
the following matricial representation of the generating function of
 the TP probability distribution:
\begin{equation}
\label{P}
\widetilde{P}^{(tr)}({\bf k};\xi)=\frac{1}{1-\xi}\left(1+{\cal D}^{-1}({\bf
k};\xi)\sum_{\mu}U_{ \mu}({\bf k};\xi)F^*({\bf
0}\;|\;{\bf e}_{\boldsymbol \mu}\;|\;{\bf Y_0};\xi)\right).
\end{equation}
In Eq.(\ref{P}) 
the function ${\cal D}({\bf k};\xi)$ stands for 
the determinant of the following $4 \times 4$ matrix,
\begin{equation}
\label{D}
{\cal D}({\bf k};\xi)\equiv{\rm det}({\rm{\bf  I-T}}({\bf k};\xi)),
\end{equation} 
where the matrix ${\rm {\bf T}}({\bf k};\xi)$ has the elements $\Big({\rm {\bf T}}({\bf k};\xi)\Big)_{\nu,\mu}$  defined by
\begin{equation}
\Big({\rm {\bf T}}({\bf k};\xi)\Big)_{\nu,\mu} = \exp\Big(i ({\bf k} \cdot {\bf
e_{\boldsymbol \nu}})\Big) \; A_{\nu,-\mu}(\xi).
\end{equation}
Explicitly, the matrix  ${\rm {\bf T}}({\bf k};\xi)$ is given by
\begin{equation}
{\rm {\bf T}}({\bf k};\xi)\equiv
\begin{pmatrix}
e^{ik_1} A_{1,-1}(\xi) & e^{ik_1} A_{1,1}(\xi) &  e^{ik_1} A_{1,-2}(\xi) &  e^{ik_1} A_{1,2}(\xi) \\
e^{-ik_1} A_{-1,-1}(\xi) & e^{-ik_1} A_{-1,1}(\xi) & e^{-ik_1} A_{-1,-2}(\xi) & e^{-ik_1} A_{-1,2}(\xi) \\
e^{ik_2} A_{2,-1}(\xi) & e^{ik_2} A_{2,1}(\xi) & e^{ik_2} A_{2,-2}(\xi) & e^{ik_2} A_{2,2}(\xi) \\
e^{-ik_2} A_{-2,-1}(\xi) & e^{-ik_2} A_{-2,1}(\xi) & e^{-ik_2} A_{-2,-2}(\xi) & e^{-ik_2} A_{-2,2}(\xi) 
\end{pmatrix},
\end{equation}
where the coefficients $A_{\nu,\mu}(\xi)$, $\nu,\mu = \pm 1, \pm 2$,  stand for
\begin{equation}
A_{\nu,\mu}(\xi) \equiv F^*({\bf 0}\;|\;{\bf e}_{\boldsymbol \nu}\;|\;{\bf e}_{\boldsymbol \mu};\xi) = \sum_{n = 0}^{+ \infty}  
F^*_n({\bf 0}\;|\;{\bf e}_{ \boldsymbol \nu}\;|\;{\bf e}_{\boldsymbol \mu}) \xi^n, 
\end{equation}
i.e. are the generating functions of the conditional probabilities for the first time visit of the origin
by the vacancy,  conditioned by constraint of the passage through a specified site on the previous step.
Note that, by symmetry,
\begin{eqnarray}
A_{2,\nu}(\xi) = A_{-2,\nu}(\xi), \nonumber\\
A_{\nu,2}(\xi) = A_{\nu,-2}(\xi)
\end{eqnarray}
for $\nu = \pm 1$ and
\begin{eqnarray}
A_{2,2}(\xi) = A_{-2,-2}(\xi), \nonumber\\
A_{2,-2}(\xi) = A_{-2,2}(\xi).
\end{eqnarray}
As a result of such a symmetry, we have to consider just ten independent functions $A_{\mu,\nu}(\xi)$ (note that in the unbiased case one
has to deal with only three such functions \cite{hilhorst1}).
Explicit expression of the determinant in Eq.(\ref{D}) in terms of these generating function is presented in the Appendix.
Lastly, the matrix $U_{{\boldsymbol \mu}}({\bf k};\xi)$ in Eq.(\ref{P}) is given by
\begin{eqnarray}
U_{\mu}({\bf k};\xi)\equiv{\cal D}({\bf k};\xi)\sum_{ \nu} 
(1-e^{-i ({\bf k} \cdot {\bf e}_{\boldsymbol \nu})})({\rm I} -{\rm T}({\bf
k};\xi))^{-1}_{\nu,\mu} \; e^{i ({\bf
k} \cdot {\bf e}_{\boldsymbol \mu})}.
\end{eqnarray}
The property of interest - the TP probability distribution function,  will be then obtained  by inverting
$\widetilde{P}^{(tr)}({\bf k};\xi)$ with respect to the wave-vector $k$ and to the variable $\xi$:
\begin{eqnarray}
P_n^{(tr)}({\bf X})=\frac{1}{2i\pi}\oint _{\cal C} \frac{{\rm
d}\xi}{\xi^{n+1}}\frac{1}{(2\pi)^2}\int_{-\pi}^{\pi}{\rm
d}k_1\int_{-\pi}^{\pi}{\rm d}k_2\;e^{i ({\bf k} \cdot {\bf X})}\widetilde{P}^{(tr)}({\bf k};\xi),
\end{eqnarray}
where the contour of integration ${\cal C}$ encircles the origin counterclockwise.

Finally, we remark that as far as we are interested in the leading large-$n$ behavior  of the
probability distribution $P_n^{(tr)}({\bf X})$ only, we may constrain 
ourselves here to the study of the asymptotic behavior of the generating function
$\widetilde{P}^{(tr)}({\bf k};\xi)$ in the vicinity of its singular point nearest to
$\xi=0$. We notice that similarly to the unbiased case, this point
is $\xi=1$ when ${\bf k}={\bf 0}$. As a matter of fact, such a behavior stems from 
the {\it a
priori} non-evident fact that the vacancy, starting from a given
neighbouring site to the origin, is certain to eventually reach the
origin. This will be demonstrated explicitly in section  4 (cf. Eq.(\ref{recurrence})); as a matter of fact, one can see from
 Eq.(\ref{recurrence}) and the explicit representation of ${\cal D}({\bf 0};\xi)$ presented in the Appendix 
 that ${\cal D}({\bf 0};\xi = 1) \equiv 0$. In consequence, expansion in powers of a small deviation $(1-\xi)$ has to
be accompanied by a small-${\bf k}$ expansion, exactly as it has been performed in Ref.\cite{hilhorst1}.

\section{The return probabilities $F^*_n({\bf 0}\;|\;{\bf
e}_{\boldsymbol \mu}\;|\;{\bf e}_{\boldsymbol \nu})$.}

As we have already remarked, the vacancy random walk between two  successive visits
of the lattice site occupied by the TP can be viewed as a standard, two-dimensional, 
symmetric random walk 
with some boundary conditions imposed 
on the four sites adjacent to the site occupied by the TP. 
In order to compute
the return probabilities  $F^*_n({\bf 0}\;|\;{\bf
e}_{\boldsymbol \mu}\;|\;{\bf e}_{\boldsymbol \nu})$ for such 
a random walk, 
we add, in a usual fashion \cite{hughes,vankampen}, an additional constraint that the site at the lattice origin is in absorbing state. Then,
the vacancy random walk can be formally represented as a lattice random walk with site-dependent probabilities of the form
 $p^+({\bf s}|{\bf s'}) = 1/4+q({\bf s}|{\bf s'})$, where ${\bf s}$ is the site occupied by the vacancy at the time moment $n$,  
${\bf s'}$ denotes the target, nearest-neighboring to ${\bf s}$ site, 
\begin{equation}
q({\bf s}|{\bf s'})\equiv
\begin{cases}
0& \text{if ${\bf s'}\notin\{{\bf 0},{\bf e_{\pm 1}},{\bf e_{\pm 2}}\}$},\\
\delta_{{\bf s},{\bf 0}}-1/4& \text{if ${\bf s'}={\bf 0}$},\\
\delta q_\nu& \text{if ${\bf s'}={\bf e}_{\boldsymbol \nu}$ and ${\bf s}={\bf
0}$},\\
-\delta q_\nu/3& \text{if ${\bf s'}={\bf e}_{\boldsymbol \nu}$ and ${\bf
s'}\neq{\bf 0}$},
\end{cases}
\end{equation} 
where $\delta q_\nu$ is defined, according to Eqs.(\ref{45}),(\ref{46}) and (\ref{zzo}), by 
\begin{equation}
\delta q_\nu\equiv\frac{p_\nu}{p_\nu+3/4}-\frac{1}{4}
\end{equation}
Further on, we define $P_{n}^+({\bf s}\;|\;{\bf s_0})$ as the probability
distribution associated with such a random walk starting at site ${\bf
s_0}$ at step $n = 0$. 

Now, let the symbols ${\cal E}$,  ${\cal A}$ and  ${\cal B}$ define the following three events:
\begin{itemize}
\item  the event ${\cal E}$: the vacancy, which has started its random walk at the site ${\bf e}_{\boldsymbol \nu}$, 
visits the
origin ${\bf 0}$ for the first time at the
$n$-th step exactly, being at  the site ${\bf e}_{\boldsymbol \mu}$ at the previous step $n-1$;
\item  the event ${\cal A}$: the vacancy, which started its random walk at the site  ${\bf e}_{\boldsymbol \nu}$,
is at the site ${\bf e}_{\boldsymbol \mu}$
at the time moment $n-1$ and the origin ${\bf 0}$ has not been visited during the $n-1$ first
steps of its walk;
\item  the event ${\cal B}$: the vacancy jumps from the neighboring to the origin 
site ${\bf e}_{\boldsymbol \mu}$
to the site ${\bf 0}$ at the $n$-th step exactly. 
\end{itemize}

Evidently,  by definition, the desired first visit probability $F^*_n({\bf 0}\;|\;{\bf e}_{\boldsymbol \mu}\;|\;{\bf e}_{\boldsymbol \nu})$
is just  the probability of the ${\cal E}$ event 
\begin{equation}
\label{99}
F^*_n({\bf 0}\;|\;{\bf e}_{\boldsymbol \mu}\;|\;{\bf e}_{\boldsymbol \nu}) = {\rm Prob}({\cal E}).
\end{equation}
To calculate ${\rm Prob}({\cal E})$ we note first that 
the probabilities of such three events obey:
\begin{equation}
\label{z}
{\rm Prob}({\cal E})={\rm Prob}({\cal A}\cap{\cal B})={\rm Prob}({\cal A})\;{\rm Prob}({\cal B}).
\end{equation}
On the other hand, we have that
\begin{equation}
\label{zz}
{\rm Prob}({\cal A})=P_{n-1}^+({\bf
e}_{\boldsymbol \mu}\;|\;{\bf e}_{\boldsymbol \nu}), 
\end{equation}
and
\begin{equation}
\label{zz1}
{\rm
Prob}({\cal B})=\frac{p_{\mu}}{3/4 + p_{\mu}}.
\end{equation} 
Hence, in virtue of Eqs.(\ref{99}),(\ref{z}),(\ref{zz}) and (\ref{zz1}), the return probability 
$F^*_n({\bf 0}\;|\;{\bf e}_{\boldsymbol \mu}\;|\;{\bf e}_{\boldsymbol \nu})$ is given explicitly by
\begin{equation}
F^*({\bf 0}\;|\;{\bf e}_{\boldsymbol \mu}\;|\;{\bf e}_{\boldsymbol
\nu};\xi)=\xi \left(\frac{p_{\mu}}{3/4 + p_{\mu}}\right) P^+({\bf
e}_{\boldsymbol \mu}\;|\;{\bf e}_{\boldsymbol \nu};\xi).
\end{equation}
Therefore, calculation of the return
probabilities  $F^*_n({\bf 0}\;|\;{\bf
e}_{\boldsymbol \mu}\;|\;{\bf e}_{\boldsymbol \nu})$ amounts to the
evaluation of the probability distribution $P_{n}^+({\bf s}\;|\;{\bf s_0})$  of the vacancy random
 walk in the presence of an
absorbing site placed at the lattice origin.
Such a probability distribution will be determined in the next subsection.

\subsection{The generating function of the probability distribution  $P^+({\bf s}\;|\;{\bf
s_0})$.}

Making use of the generating function technique adapted to random walks on lattices with defective sites 
 \cite{hughes}  and \cite{montroll}, we obtain
\begin{equation}
P^+({\bf s_i}\;|\;{\bf s_j};\xi)=P({\bf s_i}\;|\;{\bf s_j};\xi) +
\sum_{l=-2}^2 A({\bf s_i}\;|\;{\bf s_l};\xi)P^+({\bf s_l}\;|\;{\bf
s_j};\xi),
\label{pasmatricielle}
\end{equation} 
where
\begin{equation}
{\bf s_i}\equiv
\begin{cases}
{\bf e_i},& \text{for $i\in\{\pm1,\pm2\}$},\\
{\bf 0},& \text{for $i=0$},
\end{cases}
\end{equation}
and
\begin{equation}
A({\bf s_i}\;|\;{\bf s_l};\xi)\equiv
\xi\sum_{{\bf s'}} P({\bf s_i}\;|\;{\bf s'};\xi)
q({\bf s'}\;|\;{\bf s_l}),
\end{equation}
$ P({\bf s_i}\;|\;{\bf s_j};\xi)$ being the generating
function of the unperturbed associated random walk (that is,
symmetric random walk 
with no defective sites).

Further on, Eq. (\ref{pasmatricielle}) can be recast into the following matricial form:
\begin{equation}
{\rm {\bf P^+}}=({\rm {\bf 1-A}})^{-1}{\rm {\bf P}},
\label{matricielle}
\end{equation} 
in which equation ${\rm {\bf P,\;P^+,\;A}}$ stand for the $5\times5$ matrices with the elements
defined by
\begin{equation}
{\rm {\bf P}}_{i,j}=P({\bf s_i}\;|\;{\bf s_j};\xi),\;\;\;\;\;\;{\rm {\bf P^+}}_{i,j}=
P^+({\bf s_i}\;|\;{\bf s_j};\xi),\;\;\;\;\;{\rm {\bf A}}_{i,j}=A({\bf s_i}\;|\;{\bf s_j};\xi),
\end{equation}
where $i,j = 0,+1,-1,+2,-2$. 
Using next an evident relation \cite{hughes}:
\begin{equation}
\label{h}
 P({\bf s_k}\;|\;{\bf s_l};\xi) = \delta_{k,l} + \frac{\xi}{4} \sum_{\nu}   P({\bf s_k}\;|\;{\bf s_l} + {\bf e}_{\boldsymbol \nu};\xi),
\end{equation}
and the symmetry properties of a standard random walk, one can readily show that:
\begin{itemize}
\item for ${\bf s_l}\neq{\bf s_0}$ and ${\bf s_k}\neq{\bf s_0}$,
\begin{equation}
A({\bf s_k}\;|\;{\bf
s_l};\xi)=\frac{4}{3}\; \delta q_l\Big(P({\bf 0}\;|\;{\bf 0};\xi)-1-P({\bf s_k}\;|\;{\bf
s_l};\xi)+\delta_{l,k}\Big),
\end{equation}
\item  for ${\bf s_l}\neq{\bf s_0}$ and ${\bf s_k}={\bf s_0}$,
\begin{equation}
A({\bf s_0}\;|\;{\bf
s_l};\xi)=\frac{4}{3}\xi \delta q_l\left(P({\bf 0}\;|\;{\bf 0};\xi)-\frac{1}{\xi^2}(P({\bf 0}\;|\;{\bf 0};\xi)-1)\right),
\end{equation}
\item  for ${\bf s_l}={\bf s_0}$,
\begin{equation}
A({\bf s_k}\;|\;{\bf
s_0};\xi)=\delta_{k,0}-(1-\xi)P({\bf s_k}\;|\;{\bf 0};\xi),
\end{equation}
\end{itemize}
Consequently, the matrices ${\rm {\bf A}}$ and ${\rm {\bf P}}$ in Eq.(\ref{matricielle}) are given by
\begin{equation}
\label{AA}
{\rm {\bf A}}=
\begin{pmatrix}
a& \delta q_1 f& \delta q_{-1} f& \delta q_2 f& \delta q_2 f& \\
b& 0    & \delta q_{-1} e& \delta q_2 c& \delta q_2 c& \\
b& \delta q_1 e& 0      & \delta q_2 c& \delta q_2 c& \\
b& \delta q_1 c& \delta q_{-1} c& 0    & \delta q_2 e& \\
b& \delta q_1 c& \delta q_{-1} c& \delta q_2 e& 0,
\end{pmatrix},
\end{equation}
where 
\begin{eqnarray}
\label{aa}
a&\equiv&1-(1-\xi)G(\xi),\;\;\;\;b\equiv\frac{1-\xi}{\xi}(1-G(\xi)),\;\;\;\;e\equiv\frac{4}{3}(2g(\xi)-1),\nonumber\\
c&\equiv&\frac{4}{3}\left(-1+\frac{2}{\xi^2}+2G(\xi)\left(1-\frac{1}{\xi^2}\right)-g(\xi)\right),
\end{eqnarray}
and 
\begin{equation}
\label{PP}
{\rm {\bf P}}=
\begin{pmatrix}
G(\xi)& (G(\xi)-1)/\xi& (G(\xi)-1)/\xi& (G(\xi)-1)/\xi&(G(\xi)-1)/\xi\\
(G(\xi)-1)/\xi& G(\xi)& G(\xi)-2g(\xi)& \tau(\xi)& \tau(\xi)\\
(G(\xi)-1)/\xi&  G(\xi)-2g(\xi)& G(\xi)& \tau(\xi)& \tau(\xi)\\
(G(\xi)-1)/\xi&  \tau(\xi)& \tau(\xi)& G(\xi)& G(\xi)-2g(\xi)\\
(G(\xi)-1)/\xi&  \tau(\xi)& \tau(\xi)&  G(\xi)-2g(\xi)& G(\xi)
\end{pmatrix},
\end{equation}
with
\begin{eqnarray}
G(\xi)&\equiv& P({\bf 0}\;|\;{\bf 0};\xi),\;\;\;g(\xi)\equiv\frac{1}{2}\left(P({\bf
e_1}\;|\;{\bf -e_1};\xi)-P({\bf 0}\;|\;{\bf
0};\xi)\right),\;\;\; \nonumber\\
\tau(\xi)&\equiv& \left(\frac{2}{\xi^2}-1\right)G(\xi)-\frac{2}{\xi^2}+g(\xi).
\end{eqnarray}
Note that Eqs.(\ref{AA}) and (\ref{PP}) now define the $P^+$ matrix explicitly, and hence, define the generating function
of the probability
distribution $P^+({\bf s}\;|\;{\bf
s_0})$.

\subsection{Asymptotic behavior of the generating functions of the 
return probabilities in the vicinity of $\xi=1$.}

As we have already remarked, here we constrain our consideration to the analysis of the leading in $n$ behavior; this amounts to
consideration of the leading in the limit $\xi \to 1^{-}$ behavior of the corresponding generating functions.
Expanding $G(\xi)$ and $g(\xi)$ in the vicinity of the singular point $\xi = 1$, (cf
Refs.\cite{hughes} and \cite{hilhorst1,mcrea,spitzer}), we have
\begin{eqnarray}
\label{GG}
G(\xi)=\frac{1}{\pi}\ln{\frac{8}{1-\xi}}-\frac{1}{2\pi}(1-\xi)\ln(1-\xi)+{\mathcal
O}\left(1-\xi\right),\;\;\xi\to1^-,
\end{eqnarray}
and
\begin{eqnarray}
g(\xi)=\left(2-\frac{4}{\pi}\right)+\frac{2}{\pi}(1-\xi)\ln(1-\xi)+{\mathcal O}\left((1-\xi)\right),\;\;\xi\to1^-.
\end{eqnarray}
Consequently, we find by solving the matricial equation (\ref{matricielle}), that the generating 
functions of the return probabilities
obey
\begin{equation}
\label{mm}
A_{\nu,\mu}(\xi)=\frac{A_{\nu,\mu}^{(1)}(u)}{S(u)}-\frac{A_{\nu,\mu}^{(2)}(u)}{S^2(u)}\;\Big(\ln{(1-\xi)}\Big)^{-1}+{\mathcal O}\left(1-\xi\right),
\end{equation} 
where $u\equiv\exp(\beta E/2)$, 
$A_{\nu,\mu}^{(1)}(u)$ and
$A_{\nu,\mu}^{(2)}(u)$ are some rational fractions (all listed explicitly in the Appendix), while
\begin{eqnarray}
&&S(u)\equiv \Big\{\left (\pi -2\right ){u}^{6}+\left (2\,{\pi }^{2}-6\,\pi +12\right ){u}^{5}+
\left (8\,{\pi }^{2}-25\,\pi+34 \right ){u}^{4} -\nonumber\\
&-&\left (4
\,{\pi }^{2}-60\,\pi +88\right ){u}^{3} + 
\left (8\,{\pi }^{2}-25\,\pi+34 \right ){u
}^{2}+\Big(2\,{\pi }^{2}-6\,\pi +12\Big)u + \pi - 2\Big\}.
\end{eqnarray}
It follows from Eqs.(\ref{mm}) and explicit expressions for $A_{\nu,\mu}(\xi)$ presented in the Appendix,
that, in particular, the generating functions of the return probabilities fulfil:
\begin{eqnarray}
A_{1,-1}(1^-)+A_{-1,-1}(1^-)+2A_{2,-1}(1^-)&=&1\nonumber\\
A_{1,1}(1^-)+A_{-1,1}(1^-)+2A_{2,1}(1^-)&=&1\nonumber\\
A_{1,2}(1^-)+A_{-1,2}(1^-)+A_{-2,2}(1^-)+A_{2,2}(1^-)&=&1,
\label{recurrence}
\end{eqnarray}
which relations imply that the vacancy, starting its random walk from a given,
neighbouring to the origin site, is $certain$ to return eventually to the origin.

\section{The TP mean displacement and the probability distribution.}

In this section, we proceed as follows: Taking advantage of the asymptotical expansion obtained in the previous section, we
first determine the small $(1-\xi)$ behavior of the generating function   $\widetilde{P}^{(tr)}({\bf k};\xi)$, accompanied by the
small-${\bf k}$ expansion. Next, we evaluate the
generating function of the TP mean displacement, by
differentiating the obtained asymptotical expression for  $\widetilde{P}^{(tr)}({\bf k};\xi)$
with respect to the components of the wave-vector, and analyse its large-$n$ behavior. 
Lastly, we invert the asymptotical expansion  of the generating function   $\widetilde{P}^{(tr)}({\bf k};\xi)$ and obtain
the corresponding probability distribution $P^{(tr)}_n({\bf X})$ in a certain scaling limit.

\subsection{Asymptotic expansion  of the generating function $\widetilde{P}^{(tr)}({\bf k};\xi)$.}

Using the explicit representation of the determinant ${\cal D}({\bf k};\xi)$ in Eq.(\ref{D}) 
in terms of the generating functions of the
return probabilities $A_{\nu,\mu}(\xi)$, presented in the Appendix, as well as the asymptotical expansions in Eq.(\ref{mm}), we
find that  in the vicinity of $\xi =1$ and for small values of the wave-vector $k$,  ${\cal D}({\bf k};\xi)$ is given by
\begin{eqnarray}
{\cal D}({\bf k};\xi)  = i{\cal F}_1(u)k_1+{\cal F}_2(u)k_1^2+{\cal F}_3(u)k_2^2-{\cal F}_4(u)
\ln^{-1}{(1-\xi)} + \ldots , 
\end{eqnarray}
where we have used the shortenings
\begin{eqnarray}
{\cal F}_1(u)\equiv-\frac{(\pi -2)(u-1)(1+u)^{5}
(u^{2}+ 2 (2 \pi - 3) \,u +1\Big)}{
(u^2 + 2 (\pi - 1) u + 1) S(u)},
\end{eqnarray}
\begin{eqnarray}
{\cal F}_2(u)\equiv\frac{(\pi -2) (1+u)^{4} (u^{2}+1)
(u^{2}+2 (2 \pi  - 3) u+1)}{2 (u^2 + 2 (\pi - 1) u + 1) S(u)},
\end{eqnarray}
\begin{eqnarray}
{\cal F}_3(u)\equiv\frac{u (\pi -2) (1+u)^{4} ((2\pi - 3) u^{2} 
+ 2\,u+2\,\pi -3)}{(u^2 + 2 ( \pi - 1) u + 1) S(u)},
\end{eqnarray}
and 
\begin{eqnarray}
{\cal F}_4(u)\equiv\frac{\pi (\pi -2) (1+u)^{4} ((2\,\pi -3)\,{u
}^{2}+2\,u+2\,\pi -3) (u^{2}+ 2 (2 \pi -3 ) \,u + 1)}{(u^2 + 2 (\pi - 1) u + 1) S(u) },
\end{eqnarray}
and assumed, for simplicity, 
 that the starting point ${\bf Y}_0$ of the vacancy random walk is ${\bf Y}_0 = {\bf e_{-1}}$. 
On the other hand, we find that 
\begin{eqnarray}
\sum_{{\boldsymbol \nu}}U_{{\boldsymbol \nu}}({\bf k};\xi)F^*({\bf
0}\;|\;{\bf e}_{\boldsymbol \nu}\;|\;{\bf -e_1};\xi)=-i{\cal F}_1(u)k_1-{\cal F}_2(u)k_1^2-{\cal F}_3(u)k_2^2 + \ldots.
\end{eqnarray}
Consequently, in the small-${\bf k}$ limit and $\xi \to 1^-$,  the generating function $\widetilde{P}^{(tr)}({\bf k};\xi)$ obeys
\begin{eqnarray}
\label{V}
\widetilde{P}^{(tr)}({\bf
k};\xi)=\frac{1}{1-\xi}\Big\{1-\left(-i \alpha_0 k_1 +
\frac{1}{2} \alpha_1 k_1^2+\frac{1}{2}\alpha_2 k_2^2\right)\ln{(1-\xi)}\Big\}^{-1},
\end{eqnarray}
where the coefficients
\begin{equation}
\label{alpha}
\begin{cases}
\alpha_0(E)\equiv \pi^{-1} \sinh(\beta E/2)((2\pi-3)\cosh(\beta
E/2)+1)^{-1},\\
\alpha_1(E)\equiv \pi^{-1} \cosh(\beta E/2)((2\pi-3)\cosh(\beta
E/2)+1)^{-1},\\
\alpha_2(E)\equiv \pi^{-1} (\cosh(\beta
E/2)+2\pi-3)^{-1},
\end{cases}
\end{equation} 
are all functions of the field strength $E$ and of the temperature only. 

\subsection{The TP mean displacement for arbitrary field strength $E$.}

As a matter of fact, the leading large-$n$ asymptotical behavior of the TP mean displacement can be obtained directly 
from Eq.(\ref{V}), since
the generating function of
the TP mean displacement, i.e.
\begin{equation} 
\overline{{\bf X}}(\xi) \equiv \sum_{n =0}^{+\infty} \overline{{\bf X}}_n \xi^n,
\end{equation}
obeys (see, e.g. Ref.\cite{kehr}):  
\begin{eqnarray}
\label{LL}
\overline{{\bf X}}(\xi)=-i\left(\,\frac{\partial\widetilde{P}^{(tr)} }{\partial k_1} ({\bf
0};\xi) \,{\bf e_1}\,+\,\frac{\partial\widetilde{P}^{(tr)} }{\partial k_2} ({\bf
0};\xi) \,{\bf e_2}\,\right).
\end{eqnarray}
Consequently, 
differentiating  the expression on the right-hand-side side of  Eq.(\ref{V}) with respect to the components of
the
wave-vector ${\bf k}$, 
we find that the asymptotical behavior of the generating function of
the TP mean displacement in the vicinity of $\xi = 1^-$ follows
\begin{eqnarray}
\overline{{\bf X}}(\xi)\sim \Big(\frac{\alpha_0(E)}{1-\xi}\,\ln{\frac{1}{1-\xi}}\Big) {\bf e_1},
\end{eqnarray}
Further on, using the discrete Tauberian theorem
(cf. Ref.\cite{hughes}) and Eq.(\ref{alpha}), we find the following general 
force-velocity relation for the system under study
\begin{eqnarray}
\overline{{\bf X}}_n \sim  \Big(\alpha_0(E) \,\ln{n}\Big) \;{\bf e_1} =  \Big(\frac{1}{\pi} \frac{\sinh(\beta
E/2)}{(2\pi-3)\cosh(\beta
E/2)+1} \,\ln{n}\Big) \;{\bf e_1}, \;\;\; \text{as $n \to \infty$},
\label{vitesse}
\end{eqnarray}
which shows that the TP mean displacement 
grows $logarithmically$ with $n$.  In consequence, one may claim that the typical displacement along the $x_1$
direction scales as $\ln(n)$ as $n \to \infty$. 
On the other hand, 
typical displacement in the $x_2$-direction is expected 
to grow only in proportion to $\sqrt{\ln(n)}$, as in
the unbiased case \cite{hilhorst1}. These claims will be confirmed in what follows by the form of the scaling variables involved in the limiting
distribution.

Consider next behavior of the coefficient $\alpha_0(E)$ in the limit $E \to 0$. Here, we find from Eq.(\ref{alpha}) 
that
\begin{equation}
\alpha_0(E) = \frac{\beta E}{4 \pi (\pi - 1)} + {\mathcal O}(E^3),
\end{equation}
and hence, the mobility $\mu_n$, defined in Eq.(\ref{mob}), follows
\begin{equation}
\label{mob2}
\mu_n \sim \frac{\beta}{4 \pi (\pi - 1)} \; \frac{\ln{(n)}}{n}, \;\;\; \text{as $n \to \infty$}
\end{equation}
Comparing next the result in Eq.(\ref{mob2}) 
with that for the diffusivity $D_n$, Eq.(\ref{Di}), 
derived by Brummelhuis and
Hilhorst \cite{hilhorst1} in the unbiased case, we infer that the TP mobility and diffusivity do obey, at least in the leading
in $n$ order, the generalized Einstein relation of the form $\mu_n = \beta D_n$ \cite{lebowitz2}. Note, that this can not be, 
of course,
an  $\it a \; priori$ expected result, in view of an intricate nature of the random walks involved and anomalous, 
$logarithmic$ confinement of the random walk trajectories.

\subsection{Probability distribution $P^{(tr)}_n({\bf X})$.}

We turn next to calculation of the asymptotic 
forms of the probability distribution $P^{(tr)}_n({\bf X})$. Inverting 
$\widetilde{P}^{(tr)}({\bf
k};\xi)$ with respect to ${\bf k}$, we   notice first that in the limit $\xi \to 1^-$ the integrand is sharply peaked
around ${\bf k} = 0$, such that the bulk contribution to the integral 
comes from the values $k_1=0$ and $k_2=0$; this implies that  we can
extend the limits of integration from $\pm\pi$ to $\pm\infty$, which yields
in the limit $\xi\to 1^-$:
\begin{eqnarray}
\label{fur}
P^{(tr)}({\bf
X};\xi)&\sim&\frac{1}{(1-\xi) (2\pi)^2 }\,\int_{-\infty}^{+\infty} 
\int_{-\infty}^{+\infty}{\rm d}k_1{\rm d}k_2\,\exp\Big(-ik_1x_1-ik_2x_2\Big) \times \nonumber\\
&\times&
\Big\{1-\left(- i \alpha_0(E) k_1+\frac{1}{2} \alpha_1(E) k_1^2 +\frac{1}{2} \alpha_2(E) 
k_2^2\right)\ln{(1-\xi)}\Big\}^{-1}
\end{eqnarray} 
Further on, using the integral equality
\begin{eqnarray}
&&\Big\{1-\left(- i \alpha_0(E) k_1+\frac{1}{2} \alpha_1(E) k_1^2 +\frac{1}{2} \alpha_2(E) 
k_2^2\right)\ln{(1-\xi)}\Big\}^{-1} =\nonumber\\
&&\int_0^{+\infty}{\rm
d}v\, \exp\Big(- v \Big\{1-\left(- i \alpha_0(E) k_1+\frac{1}{2} \alpha_1(E) k_1^2 +\frac{1}{2} \alpha_2(E) 
k_2^2\right)\ln{(1-\xi)}\Big\}\Big)
\end{eqnarray}
we cast the integral in Eq.(\ref{fur}) into the form:
\begin{eqnarray}
\label{intgaussienne}
&&P^{(tr)}({\bf
X};\xi)\sim\frac{1}{(1 - \xi) (2\pi)^2}  \int_0^{+\infty}{\rm d}v \exp(-v) \int_{-\infty}^{+\infty}{\rm
d}k_1\int_{-\infty}^{+\infty}{\rm
d}k_2 \exp(-i k_2 x_2) \times \nonumber\\
&\times& \exp\Big(\frac{v}{2}\left( \alpha_1(E) k_1^2+ \alpha_2(E) k_2^2\right)\ln{(1-\xi)}  
-ik_1\left(x_1 + v \alpha_0(E) \ln{(1-\xi)}\right) \Big)
\end{eqnarray}
Note now that in order to evaluate explicitly the Gaussian integral in Eq.(\ref{intgaussienne}), we have 
to consider separately two cases: when (a) the external field in infinitely strong, $E =
\infty$ (which implies $\alpha_2 = 0$), such that the TP performs a totally directed walk, and (b) - when $E$ is bounded, $E <
\infty$ (and hence, $\alpha_2 > 0$). 

\subsubsection{Directed walk, $E =
\infty$.}

We start with the simplest case when the TP 
performs a totally directed walk under the 
influence of an infinitely strong
field.  In this case, the probability
distribution is defined for  non negative $x_1$ values only, and  the equation
(\ref{intgaussienne}) reduces to:
\begin{eqnarray}
P^{(tr)}({\bf
X};\xi)&\sim&\frac{\delta(x_2) \; \theta(x_1)}{2\pi (1-\xi)}
\int_0^{+\infty}{\rm d}v \exp(-v) \int_{-\infty}^{+\infty}{\rm
d}k_1   \times \nonumber\\
&\times&  \exp\Big( \frac{v}{2} \alpha_1(E) k_1^2\ln{(1-\xi)}
-ik_1\left(x_1+v \alpha_0(E) \ln{(1-\xi)}\right)
\Big)
\end{eqnarray}
where $\theta(x_1)$ denotes the Heaviside theta-function. 
Performing the integrals, we find that,  in the limit
$\xi\to1^-$,  the generating function $P^{(tr)}({\bf
X};\xi)$ obeys:
\begin{eqnarray}
P^{(tr)}({\bf
X};\xi)\sim - \delta(x_2)\theta(x_1)\frac{\pi(2\pi-3) }{(1-\xi)\ln{(1-\xi)}} 
\exp\Big(\frac{\pi(2\pi-3)}{\ln{(1-\xi)}} x_1
\Big).
\end{eqnarray}
Applying next the discrete Tauberian theorem \cite{hughes,feller}, we find eventually,
\begin{eqnarray}
P^{(tr)}_n({\bf
X})\sim\delta(x_2)\theta(x_1)\frac{\pi(2\pi-3)}{\ln{(n)}} \exp\Big(-\frac{\pi(2\pi-3)}{\ln{(n)}} x_1\Big),
\end{eqnarray}
which means that in the totally directed case, in the large-$n$ and large-$x_1$ limit,  the
scaled variable $\eta_{\infty} \equiv \pi (2\pi - 3) x_1/\ln{(n)}$ is asymptotically
 distributed according to 
\begin{equation}
P(\eta_{\infty}) = \theta(\eta_{\infty}) \exp(- \eta_{\infty}), 
\end{equation}
i.e. has an exponential scaling function. 

\subsubsection{Arbitrary bounded field $E <
\infty$.}

In this case the coefficient $\alpha_2 > 0$ and the probability distribution is defined also for negative values of
$x_1$; as well, $P^{(tr)}({\bf
X};\xi)$ is defined also for non-zero values of $x_2$. In this general case, we find, performing integrations over the
components of the wave-vector, that $P^{(tr)}({\bf
X};\xi)$ attains, as $\xi\to1^-$, the following form:
\begin{eqnarray}
&&P^{(tr)}({\bf
X};\xi) \sim - \Big(2\pi (1-\xi)\ln{(1-\xi)} \sqrt{\alpha_1(E) \alpha_2(E) }\Big)^{-1} \int_0^{+\infty}{\rm d}v  \exp( -
v) \times \nonumber\\
&\times& \exp\Big(\frac{1}{2v \ln{(1-\xi)}} 
\left(\left(\frac{x_1}{\alpha_1(E)}+ v \frac{\alpha_0(E)}{\alpha_1(E)}\ln{(1-\xi)}\right)^2+
\left(\frac{x_2}{\alpha_2(E)}\right)^2\right)
\Big)
\end{eqnarray}
The integral in the latter equation can be calculated exactly, which yields  
\begin{eqnarray}
P^{(tr)}({\bf
X};\xi)\sim  - \Big(\pi (1-\xi)\ln{(1-\xi)} \sqrt{\alpha_1(E) \alpha_2(E) }\Big)^{-1} \; 
\exp\Big(\frac{\alpha_0(E)}{\alpha_1(E)} x_1\Big) \; K_0\left(\eta_E \left(\frac{1}{1-\xi}\right)\right),
\end{eqnarray}
where $K_0$ is the modified Bessel (McDonald) function of zeroth order, and
\begin{equation}
\label{KK}
\eta_E(\lambda) \equiv \sqrt{\frac{2}{\ln{(\lambda)}}+\frac{\alpha_0^2(E)}{\alpha_1(E)}}
\sqrt{\frac{x_1^2}{\alpha_1(E)}+\frac{x_2^2}{\alpha_2(E)}}
\end{equation}
Finally, using the discrete Tauberian theorem \cite{hughes,feller}, we find from Eq.(\ref{KK}) that in the 
large-$n$ and large-$X$ limits, the probability distribution $P^{(tr)}_n({\bf
X})$ obeys
\begin{eqnarray}
\label{U}
P^{(tr)}_n({\bf
X})\sim \Big(\pi \sqrt{\alpha_1(E) \alpha_2(E)} \ln{(n)} \Big)^{-1} \; \exp\Big(\frac{\alpha_0(E)}{\alpha_1(E)} x_1\Big) \; K_0(\eta_E(n)).
\end{eqnarray}
Note that in the unbiased case, i.e. when $E = 0$, the probability distribution $P^{(tr)}_n({\bf
X})$ defined by Eqs.(\ref{KK}) and (\ref{U}) reduces to the form predicted earlier by Brummelhuis and Hilhorst \cite{hilhorst1}.

\subsubsection{Limiting probability distribution function.}

Now, 
we recollect that the scaling behavior 
expected is $x_1 \sim \ln{(n)}$ (for $E >  0$)and $x_2 \sim \sqrt{\ln{(n)}}$. In order to obtain from 
Eqs.(\ref{KK}) and (\ref{U})   the
limiting probability distribution, we introduce two scaling variables:
\begin{equation}
\label{al}
\begin{cases}
\eta_1\equiv x_1/\alpha_0(E) \ln{(n)},\\
\eta_2\equiv x_2/\sqrt{2 \alpha_2(E) \ln{(n)}}.
\end{cases}
\end{equation} 
Note that $\eta_1$ becomes $\eta_{\infty}$ in the special case $E = \infty$. 
In terms of these scaling variables $\eta_E(n)$ in Eq.(\ref{KK}) takes the form:
\begin{equation}
\eta_E(n) = \frac{\alpha_0^2(E)}{\alpha_1(E)} \; \ln{(n)} \; |\eta_1| \; \Big(1 + \frac{\alpha_1(E)}{\alpha_0^2(E) \ln{(n)}} 
\Big(1 + (\frac{\eta_2}{\eta_1})^2\Big)  + {\mathcal O}(1/\ln^2{(n)}) \Big)
\end{equation}
Note now that for arbitrary fixed $\eta_1$ and $\eta_2$, 
the argument of the Bessel function
$\eta_E(n)$ written in terms of the scaling variables tends to infinity as $n \to \infty$.
Consequently, using the limiting behavior of the modified Bessel
function
\begin{equation}
K_0(y) = \Big(\frac{\pi}{2 y}\Big)^{1/2} \exp( - y) (1 + {\mathcal O}(1/y)),
\end{equation}
we find that the probability distribution $P^{(tr)}_n({\bf
X})$, written in terms of the scaling variables, converges as $n \to \infty$ to the limiting form
\begin{eqnarray}
P^{(tr)}_n({\bf
X}) \sim_{n \to \infty}  
\begin{cases}
(2 \pi \alpha_2(E) \alpha_0^2(E) \eta_1 \ln^3{(n)})^{-1/2} \exp(- \eta_1 - \eta_2^2/\eta_1), \;\;\; \text{for $\eta_1 \geq 0$},\\
0,  \;\;\; \text{for $\eta_1 < 0$},
\end{cases}
\end{eqnarray}
or, equivalently, that the 
scaling variables $\eta_1$ and $\eta_2$ have  the following, rather unusual
limiting joint distribution function:
\begin{equation}
 P(\eta_1,\eta_2) = \frac{\theta(\eta_1)}{\sqrt{\pi \eta_1}} \exp\Big(- \eta_1 - \frac{\eta_2^2}{\eta_1}\Big)
\end{equation}
We note that this distribution is properly normalized and yields, of course, the same result for the TP mean displacement as the
approach based on differentiation of the asymptotical expansion of the generating function. 
We also remark that the reduced distributions $P(\eta_1) = \int d\eta_2  P(\eta_1,\eta_2)$ and $P(\eta_2) 
= \int d\eta_1  P(\eta_1,\eta_2)$ take the form:
\begin{eqnarray}
P(\eta_1) &=& \theta(\eta_1) \exp( - \eta_1), \nonumber\\
P(\eta_2) &=&  \exp( - 2 |\eta_2|),
\end{eqnarray}
and hence, the reduced distribution $P(\eta_1)$ appears to be exactly the same as in the case $E = \infty$.

\section{Finite vacancy concentration.}

In this last section, we generalize our analysis of the biased TP mean displacement to the case when 
vacancies are present at a very small, but finite concentration $\rho$.  
In our approach, we follow closely that of Brummelhuis and Hilhorst \cite{hilhorst2}, 
who pointed out that for the unbiased case in the limit of low vacancy concentration, the
many-vacancy problem can be interpreted in terms of
 the one-vacancy solution, which entails meaningful results to the leading order in the concentration of
vacancies for $\rho \ll 1$. We thus just extend here their consideration over the biased case.
 
Following Ref.\cite{hilhorst2}, we begin by considering a finite lattice of size $L\times L$, containing $M$ 
vacancies. The mean concentration of the vacancies is thus $\rho = M/L^2 \ll 1$. 
We suppose that the charged TP is initially at the origin and initial positions of the vacancies are ${\bf Y}^{(1)}_0, {\bf Y}^{(2)}_0, \ldots ,{\bf Y}^{(M)}_0$, 
which all are different from each other and from 
${\bf 0}$. All other sites are filled with neutral hard-core particles. 
The field ${\bf E}$ is again supposed to be oriented in the positive $x_1$-direction.

Similar to the single vacancy case, we stipulate that at each time step, all vacancies exchange their positions with either of neighboring particles, 
such that each
vacancy makes a step each time step. Exchanges with the charged TP are governed by the same rules as described in Section 2. 
Note that, of course, when many vacancies are present, 
it may appear that two or more vacancies occupy adjacent sites or have common neighboring particles, 
in which case their random walks will interfere. However, as noticed in \cite{hilhorst2}, these cases contribute only to $\mathcal{O}(\rho^2)$ and thus go
 beyond our approximation; hence, we discard such possibilities here.      

Now, let ${\cal P}_n^{(tr)}({\bf X}|{\bf Y}^{(1)}_0, {\bf Y}^{(2)}_0, \ldots ,{\bf Y}^{(M)}_0)$ denote the
 probability of finding at time moment $n$ the TP 
at position ${\bf X}$  as a result of its interaction with all $M$ vacancies collectively. Further on, 
let $P_n^{(tr)}({\bf X}^{(j)}|{\bf Y}^{(j)}_0)$  denote 
the probability of finding the TP at site ${\bf X}^{(j)}$ at time moment $n$ due to interactions with a vacancy initially 
at ${\bf Y}^{(j)}_0$ in a system with a $single$
vacancy. Then, assuming that the vacancies contribute independently to the TP displacement, one has, following 
Ref.\cite{hilhorst2}:
\begin{eqnarray}
{\cal P}_n^{(tr)}( {\bf X}|{\bf Y}^{(1)}_0, {\bf Y}^{(2)}_0, \ldots ,{\bf Y}^{(M)}_0 )\approx
\sum_{{\bf Y}^{(1)}_0} \cdots \sum_{{\bf Y}^{(M)}_0} 
\delta_{{\bf X}, {\bf Y}^{(1)}_0  + \ldots + {\bf Y}^{(M)}_0 } \prod_{j = 1}^M P_n^{(tr)}({\bf X}^{(j)}|{\bf Y}^{(j)}_0)
\end{eqnarray}
Upon averaging over all initial vacancy configurations, and denoting
this average by the angular brackets, we have
\begin{eqnarray}
\Big<{\cal P}_n^{(tr)}( {\bf X}|{\bf Y}^{(1)}_0, {\bf Y}^{(2)}_0, \ldots ,{\bf Y}^{(M)}_0 )\Big>    \approx
\sum_{{\bf Y}^{(1)}_0} \cdots \sum_{{\bf Y}^{(M)}_0} 
\delta_{{\bf X}, {\bf Y}^{(1)}_0  + \ldots + {\bf Y}^{(M)}_0 } \Big<   \prod_{j = 1}^M P_n^{(tr)}({\bf X}^{(j)}|{\bf Y}^{(j)}_0) \Big>,
\end{eqnarray}
which, in the limit of very small vacancy concentration, simplifies to \cite{hilhorst2}:
\begin{eqnarray}
\Big<{\cal P}_n^{(tr)}( {\bf X}|{\bf Y}^{(1)}_0, {\bf Y}^{(2)}_0, \ldots ,{\bf Y}^{(M)}_0 )\Big>    \approx
\sum_{{\bf Y}^{(1)}_0} \cdots \sum_{{\bf Y}^{(M)}_0} 
\delta_{{\bf X}, {\bf Y}^{(1)}_0  + \ldots + {\bf Y}^{(M)}_0 }  \prod_{j = 1}^M  \Big< P_n^{(tr)}({\bf X}^{(j)}|{\bf Y}^{(j)}_0) \Big>
\end{eqnarray}

Now, defining the Fourier transformed distributions
\begin{equation}
\widetilde{{\cal P}}^{(tr)}_n({\bf
k},M,L) = \sum_X \exp(i ({\bf k } \cdot {\bf X})) \Big<{\cal P}_n^{(tr)}( {\bf X}|{\bf Y}^{(1)}_0, {\bf Y}^{(2)}_0, \ldots ,{\bf
Y}^{(M)}_0 )\Big> ,
\end{equation}
and 
\begin{equation}
\widetilde{P}^{(tr)}_n({\bf
k}) = \sum_X \exp(i ({\bf k } \cdot {\bf X})) P_n^{(tr)}( {\bf X}|{\bf Y}^{(1)}_0),
\end{equation}
and performing the corresponding summations, we find that 
\begin{eqnarray}
\widetilde{{\cal P}}^{(tr)}_n({\bf
k},M,L) \approx\left( \widetilde{P}_n^{(tr)}({\bf k})\right)^M
\end{eqnarray}
Turning next to the limit $L, M \to \infty$ (while the ratio $M/L^2 = \rho$ is kept fixed),  we obtain
\begin{eqnarray} 
\label{PPP}
\widetilde{{\cal P}}^{(tr)}_n({\bf
k},\rho) =  \lim_{L,M \to \infty}    \widetilde{{\cal P}}^{(tr)}_n({\bf
k},M,L)   \approx \exp\Big(-\rho \; \Omega_n({\bf k})\Big),
\end{eqnarray}
where
\begin{eqnarray}
\Omega_n({\bf k})\equiv\sum_{j=0}^n\sum_\nu\Delta_{n-j}({\bf k}|{\bf e}_{\boldsymbol
\nu})\sum_{{\bf Y\neq0}} F^*_j({\bf
0}|{\bf e}_{\boldsymbol \nu}|{\bf Y}),
\end{eqnarray}
\begin{eqnarray}
\Delta_n({\bf k}|{\bf e}_{\boldsymbol \nu})=1-\widetilde{P}_n^{(tr)}({\bf
k}|{\bf -e}_{\boldsymbol \nu}) \exp(i ({\bf k} \cdot {\bf e}_{\boldsymbol \nu})),
\end{eqnarray}
$ F^*_j({\bf
0}|{\bf e}_{\boldsymbol \nu}|{\bf Y})$ are conditional return probabilities,  defined in Section 3,  and $\widetilde{P}_n^{(tr)}({\bf
k}|{\bf -e}_{\boldsymbol \nu})$ is the Fourier transformed single-vacancy probability distribution $P_n^{(tr)}({\bf
X}|{\bf -e}_{\boldsymbol \nu})$. The latter can be readily obtained by applying the discrete Fourier transformation 
to the relation
\begin{eqnarray}
P_n^{(tr)}({\bf X}|{\bf Y})&=& \delta_{{\bf X},{\bf 0}}\left(1-\sum_{j=0}^n F^*_j({\bf
0}|{\bf Y})\right)+\nonumber\\ 
&+&\sum_{j=0}^n\sum_\nu P_{n-j}^{(tr)}({\bf X-e}_{\boldsymbol \nu}\;|\;{\bf -e}_{\nu}) F^*_j({\bf
0}|{\bf e}_{\boldsymbol \nu}|{\bf Y}).
\end{eqnarray} 
and chosing ${\bf Y } = {\bf -e}_{\boldsymbol \nu}$.

Further on, using the results of the previous section, we find that in the limit  $\xi\to1^-$, ${\bf k}\to{\bf 0}$, the generating function
of $\Omega_n({\bf k})$ is given by 
\begin{eqnarray}
\Omega({\bf k};\xi)=\sum_\nu\Delta({\bf k}|{\bf e}_{\boldsymbol
\nu};\xi)\sum_{{\bf Y\neq0}} F^*({\bf
0}|{\bf e}_{\boldsymbol \nu}|{\bf Y};\xi),
\end{eqnarray}
with
\begin{eqnarray}
\Delta({\bf k}|{\bf e}_{\boldsymbol \nu};\xi)&\equiv&
\frac{1}{1-\xi}\Big(1-\exp(i ({\bf k} \cdot {\bf e}_{\boldsymbol \nu}) )
\Big\{1 - \ln{(1-\xi)} \Big(-i \alpha_0(E) k_1+\nonumber\\
&+&\frac{1}{2} \alpha_1(E) k_1^2+\frac{1}{2}\alpha_2(E) k_2^2\Big)\Big\}^{-1}
\Big) + \ldots .
\end{eqnarray}
We turn next to calculation of $\sum_{{\bf Y\neq0}} F^*({\bf
0}|{\bf e}_{\boldsymbol \nu}|{\bf Y};\xi)$ in the limit $\xi\to1^-$,
${\bf k}\to{\bf 0}$, which can be done rather straightforwardly by taking advantage of the results of Section 3.  We have then
\begin{eqnarray}
\label{80}
&&\sum_{{\bf Y\neq0}} F^*({\bf
0}|{\bf e}_{\boldsymbol \nu}|{\bf Y};\xi)=\xi\left(\frac{p_{\nu}}{3/4+ p_\nu}\right)\sum_{{\bf Y\neq0}} P^+(
{\bf e}_{\boldsymbol \nu}|{\bf
Y};\xi)=\nonumber\\
&&\xi\left(\frac{p_{\nu}}{3/4+ p_\nu}\right) {\bf {\cal B}}^t_\nu({\rm {\bf
1-A}})^{-1} \sum_{{\bf Y\neq0}}{\bf {\cal B}}({\bf Y};\xi),
\end{eqnarray}
where ${\bf {\cal B}_\nu}$ is the $\nu-th$ basis vector, ${\bf {\cal B}}^t_\nu$ denotes the 
transposition of  ${\bf {\cal B}_\nu}$, 
 and ${\bf {\cal B}}({\bf
Y};\xi)$ is the vector, whose elements are $(P({\bf s_i}|{\bf Y};\xi))_i$, $i = 0,1,-1,2,-2$. Explicitly,
\begin{equation}
{\bf {\cal B}_0} \equiv
\begin{pmatrix}
1\\0\\0\\0\\0
\end{pmatrix}, \;\;
{\bf {\cal B}_1} \equiv
\begin{pmatrix}
0\\1\\0\\0\\0
\end{pmatrix}, \;\;
{\bf {\cal B}_{-1}} \equiv
\begin{pmatrix}
0\\0\\1\\0\\0
\end{pmatrix}, \;\;
{\bf {\cal B}_{2}} \equiv
\begin{pmatrix}
0\\0\\0\\1\\0
\end{pmatrix}, \;\;
{\bf {\cal B}_{-2}} \equiv
\begin{pmatrix}
0\\0\\0\\0\\1
\end{pmatrix}
\end{equation}

Further on, using an evident  symmetry relation
\begin{equation}
P(s_i | {\bf Y}_i; \xi) = P({\bf Y}_i | s_i; \xi), 
\end{equation} 
as well as the relation in Eq.(\ref{h}), we obtain
\begin{eqnarray}
\label{81}
\sum_{{\bf Y\neq0}}{\bf {\cal B}}({\bf
y};\xi)=\left(\frac{1}{1-\xi}-G(\xi)\right){\bf {\cal
B}_0}+\left(\frac{1}{1-\xi}-\frac{1}{\xi}(G(\xi)-1)\right)\left({\bf {\cal B}_1+{\cal B}_{-1}+{\cal B}_2+{\cal B}_{-2}}\right).
\end{eqnarray}
Then, combining Eqs.(\ref{80}), (\ref{81}) and (\ref{AA}), (\ref{aa}), (\ref{GG}), and performing some straightforward but cumbersome calculations, 
we find 
that in the limit $\xi\to1^-$ and ${\bf k}\to{\bf 0}$, the sum $\sum_{{\bf Y\neq0}} F^*({\bf
0}|{\bf e}_{\boldsymbol \nu}|{\bf Y};\xi)$ is given by 
\begin{eqnarray}
\sum_{{\bf Y\neq0}} F^*({\bf
0}|{\bf e}_{\boldsymbol \nu}|{\bf Y};\xi)=- \frac{\pi}{(1-\xi)\ln{(1-\xi)}}+ \ldots,
\end{eqnarray}
which is, remarkably, independent of $u$ and $\nu$ in the leading in $\xi$ order. 
Consequently, in the limit $\xi\to1^-$ and ${\bf k}\to{\bf 0}$, the generating function $\Omega({\bf k};\xi)$ obeys:
\begin{eqnarray}
\Omega({\bf k};\xi) \approx\frac{\pi}{(1-\xi)^2}\frac{-i \alpha_0(E) k_1 +\frac{1}{2}
\alpha_1(E) k_1^2+\frac{1}{2}\alpha_2(E) k_2^2}{1 - \ln{(1-\xi)}\left(-i \alpha_0(E) k_1 + 
\frac{1}{2} \alpha_1(E) k_1^2 + \frac{1}{2}\alpha_2(E) k_2^2\right)}.
\end{eqnarray}
Next, using the discrete Tauberian theorem, we obtain from the latter equation that 
in the limit $n \to \infty$ and ${\bf k}\to{\bf 0}$, 
\begin{eqnarray}
\label{O}
\Omega_n({\bf k}) \approx \pi \frac{\Big(-i \alpha_0(E) k_1 +\frac{1}{2}
\alpha_1(E) k_1^2+\frac{1}{2}\alpha_2(E) k_2^2\Big) }{1 + \ln{(n)}\left(-i \alpha_0(E) k_1 + 
\frac{1}{2} \alpha_1(E) k_1^2 + \frac{1}{2}\alpha_2(E) k_2^2\right)} \; n.
\end{eqnarray}
Finally, inverting  Eq.(\ref{PPP}) 
with respect to the wave-vector,
\begin{equation}
 {\cal P}^{(tr)}_n({\bf
X},\rho)  \approx \frac{1}{4 \pi^2} \int^{\pi}_{-\pi} dk_1 \int^{\pi}_{-\pi} dk_2 \exp\Big( - i ({\bf k} \cdot {\bf X}) - \rho \; \Omega_n({\bf k})\Big), 
\end{equation}
and taking advantage of Eq.(\ref{LL}), 
we find that the leading, large-$n$ 
behavior of the TP mean displacement is given by
\begin{eqnarray}
\overline{{\bf X}}_n \sim \Big(\pi\,\alpha_0(E) \; \rho \, n \Big)\;{\bf e_1}=\,\frac{\sinh(\beta
E/2)}{(2\pi-3)\cosh(\beta E/2)+1} \rho \,n\; {\bf e_1},
\end{eqnarray}
i.e. grows linearly with time. This signifies that the TP mobility attains a  constant value at sufficiently large times $n$,
\begin{equation}
\label{mobb}
\mu_n = \lim_{|E| \to 0}\frac{|\overline{{\bf X}}_n|}{|{\bf E}| n} = \frac{\beta \rho}{4 (\pi - 1)}
\end{equation}
Lastly, on comparing Eq.(\ref{mobb}) and the result of Brummelhuis and Hilhorst \cite{hilhorst2} for the TP diffusivity in absence of the field, Eq.(\ref{diff}), 
we notice that again the Einstein relation is fulfilled!

\section{Conclusion}

In conclusion, we have studied the dynamics of a charged tracer particle 
diffusing 
on a two-dimensional lattice, all sites of which except one (a vacancy) are 
filled with identical neutral, hard-core particles. 
The system evolves in discrete time $n$, $n = 0, 1, 2, \ldots $, by particles 
exchanging their positions with the vacancy, subject to 
the condition that each site can be at most singly occupied. 
The charged TP experiences a bias due to external 
field ${\bf E}$, which favors its jumps in the
preferential direction. 
We determine exactly, 
for arbitrary strength of the field $E = |{\bf E}|$, the leading large-$n$ 
behavior of the TP mean displacement $\overline{{\bf X}}_n$, which is not zero here due to external bias,
and the limiting probability distribution of the TP position. We have shown that
the TP trajectories are anomalously confined and its mean displacement grows with time only 
logarithmically,  $\overline{{\bf X}}_n = (\alpha_0(E) \; \ln(n)) \; {\bf e}_1$ as $n \to \infty$.
On comparing our results  with the earlier analysis of the 
TP diffusivity $D_n$ in the unbiased case by
 Brummelhuis and Hilhorst \cite{hilhorst1}, we have demonstrated that, remarkably, 
the Einstein relation $\mu_n = \beta
D_n$ between the 
diffusivity and the mobility $\mu_n$ 
of the TP holds in the leading in $n$ order, 
despite the fact that both  $D_n$ and $\mu_n$ tend to zero 
as $n \to \infty$. Note, however, that validity of the Einstein relation for the system under study 
relies heavily on the proper normalization of the vacancy
transition probabilities (see, Eqs.(\ref{45}),(\ref{46}) and (\ref{zzo})). In absence of such a normalization, 
artificial "temporal trapping" effects may emerge, which will result ultimately in the violation of the Einstein relation
for the system under study (see also Refs.\cite{cugl1} and \cite{cugl2} for physical situations in 
which such type of effects is observed).  
Further on,
we have also generalized
 our approach to the situation with small but finite vacancy concentration $\rho$, 
in which case we have
found a ballistic-type law of the form $\overline{{\bf X}}_n = ( \pi \; \alpha_0(E) \; \rho \;  
n) \; {\bf e}_1$. We have shown  that here, 
again, both  $D_n$ and  $\mu_n$
calculated in the linear in $\rho$ approximation do obey the Einstein relation.

\vspace{0.3in}
{\Large \bf Acknowledgments.}
\vspace{0.3in}

The authors wish to thank Prof. H.Hilhorst for helpful discussions. 
We are also very grateful to Dr. S. Nechaev for valuable comments on the
manuscript.  

\newpage

\section{Appendix.}

In this Appendix, we list some explicit expressions skipped in the body of the manuscript. 
First of all, explicit form of the determinant ${\cal D}({\bf k};\xi)$ in Eq.(\ref{D}) in terms of the generating functions of the
return probabilities $A_{\nu,\mu} = A_{\nu,\mu}(\xi)$ reads

\begin{eqnarray}
&&{\cal D}({\bf k};\xi)=1-A_{2,2}^2+2\,A_{-1,-1} A_{2,1} A_{1,2} A_{-2,2}+2\,
A_{1,-1} A_{2,1}  A_{-1,2} A_{2,2} -  \nonumber\\
&-&2\, A_{2,-1} A_{-2,2} A_{1,2} A_{-1,1} - 
A_{1,-1} A_{-1,1} A_{2,2}^2+ 
A_{1,-1} A_{-1,1} A_{-2,2}^2+ \nonumber\\
&+&A_{1,-1} A_{-1,1}-A_{-1,-1} A_{1,1}
-A_{-1,-1} A_{1,1} A_{-2,2}^2
+  A_{-1,-1} A_{1,1} A_{2,2}^2
-\nonumber\\
&-&
2\,A_{-1,-1}A_{2,1}A_{1,2}A_{2,2}
+2\,A_{2,-1} A_{2,2} A_{1,2} A_{-1,1} -2\,A_{1,-1} A_{2,1} A_{-1,2} A_{-2,2} -\nonumber\\
&-&2\,A_{2,-1} A_{2,2} A_{1,1} A_{-1,2} 
+2\,A_{2,-1} A_{-2,2} A_{1,1} A_{-1,2} +A_{-2,2}^2 +\nonumber\\
&+&2\,\Big(A_{2,-1} A_{1,2} A_{-1,1} +
A_{1,-1} A_{2,1} A_{-1,2} -
A_{-1,-1} A_{2,1} A_{1,2} -\nonumber\\
&-&A_{1,-1} A_{-1,1} A_{-2,2} 
-A_{-2,2} +A_{-1,-1} A_{1,1} A_{-2,2} -
A_{2,-1} A_{1,1} A_{-1,2} \Big)\cos{k_2}+\nonumber\\
&+& 
\Big(
A_{-1,1} A_{2,2}^2 -A_{-1,1} +2\,A_{2,1} A_{-1,2} A_{-2,2} -
2\,A_{2,1} A_{-1,2} A_{2,2} -\nonumber\\
&-&A_{-1,1} A_{-2,2}^2 \Big)e^{-ik_1}
+\Big(2\,A_{2,-1} A_{-2,2} A_{1,2} -2\,A_{2,-1} A_{2,2} A_{1,2} +\nonumber\\
&+&A_{1,-1} A_{2,2}^2 -A_{1,-1} -
A_{1,-1} A_{-2,2}^2 \Big)e^{ik_1}
+2\,\Big(A_{1,-1} A_{-2,2} -\nonumber\\
&-&A_{2,-1} A_{1,2} \Big)e^{ik_1}\cos{k_2}+
2\,\Big(-A_{2,1} A_{-1,2} +A_{-1,1} A_{-2,2} \Big)e^{-ik_1}\cos{k_2} \nonumber\\
\nonumber
\end{eqnarray} 

Next, the coefficients in Eq.(\ref{mm}) 
defining asymptotical behavior of the generating functions of the return probabilities are
given explicitly by:
\begin{eqnarray}
A_{1,-1}^{(1)}(u)&=&u^{2}(u+1)^{2}((\pi -2) u^{2} - 2 (\pi^2 - 3 \pi  -  2) u + \pi -2), \nonumber\\
A_{1,-1}^{(2)}(u)&=&-\pi u^{2} \,(u+1)^{2}(u^{2}+2\,u+2\,\pi -3) \times \nonumber\\
&\times& \Big((2\,\pi -3 ) u^{2} +2\,u + 1) ((\pi - 2) u^2 +4\,u +\pi - 2)^{2}, \nonumber\\
A_{-1,-1}^{(1)}(u)&=&(4 \,{\pi }^{2}-15\,\pi +14){u}^{4}-(6\,\pi^{2}-56\,\pi +80\Big){u}^{3}+\nonumber\\
&+&(8\,\pi^{2}-34\,\pi +52) u^{2}+(2\,\pi^{2}-8\,\pi +16)u+\pi -2, \nonumber\\
A_{-1,-1}^{(2)}(u)&=&-\pi \, (u+1)^{2} (( 2\,\pi-3) {u}^{2} +2\,u + 1)^{2}
((\pi - 2) u^2 + 4\,u+\pi -2)^{2},\nonumber\\
A_{2,-1}^{(1)}(u)&=& u (\pi -2)(u+1)^{2} (( 2\,\pi - 3) u^2 +2\,u + 1), \nonumber\\
A_{2,-1}^{(2)}(u)&=&-\pi \,u (u+1)^{2} ((2\,\pi -3)u^2+2\,u+1)
((\pi - 2) u^2
+4\,u+\pi-2) \times\nonumber\\
&\times& ((\pi^{2}-4 \pi + 6) {u}^{4} 
+( 2\,\pi^{2}-6\,\pi +8)
u^{3}-(2\,\pi^{2}-\nonumber\\
&-&20\,\pi + 28) \,u^{2}+(2\,\pi^{2}
-6 \pi +8)\,u+\pi^{2}-4\,\pi +6), \nonumber\\
\nonumber
\end{eqnarray}
\begin{eqnarray}
A_{1,1}^{(1)}(u)&=& u^{2}((\pi -2){u}^{4}+(2\,\pi^{2}-8\,\pi +16) u^{3}+(8\,\pi^{2}-34\,\pi +52) u^{2}-\nonumber\\
&-&(6\,{\pi }^{2}-56 \pi + 80) u+4\,{\pi }^{2}-15\,\pi+14), \nonumber\\
A_{1,1}^{(2)}(u)&=&-u^{2} \pi \,(u+1)^{2}(u^{2}+2\,u+2\,\pi -3)^{2}
((\pi - 2) u^2 + 4\,u+\pi-2)^{2},\nonumber\\
A_{-1,1}^{(1)}(u)&=&(u+1)^{2}
((\pi - 2) {u}^{2} -2 (\pi^2 - 3\,\pi -2)\,u  + \pi -2), \nonumber\\
A_{-1,1}^{(2)}(u)&=&-\pi \,(u+1)^{2}
({u}^{2}+2\,u+2\,\pi -3) \times\nonumber\\
&\times&((2\pi - 3) u^{2}+2\,u+1)((\pi -2) {u}^{2}+4\,u +\pi-2)^{2}, \nonumber\\
A_{2,1}^{(1)}(u)&=&(\pi -2) u (u+1)^{2}(u^{2}+2\,u+2\,\pi 
-3), \nonumber\\
A_{2,1}^{(2)}(u)&=&- \pi \,u (u+1)^{2}(u^{2}+2\,u+2\,\pi -3)
((\pi - 2) {u}^{2} +
4\,u +\pi 
-2) \times\nonumber\\
&\times&(({\pi }^{2}-4\,\pi+6 ){u}^{4} +(2\,{\pi }^{2} -6\pi  +8) \,{u}^{3}
-\nonumber\\
&-&(2
\,{\pi }^{2}-20\,\pi +28)\,{u}^{2}+
(2\,{\pi }^{2}-6\,
\pi +8)\,u+{\pi }^{2}-4\,\pi +6), \nonumber\\
A_{1,2}^{(1)}(u)&=&{u}^{2}(\pi -2)(u+1)^{2}({u}^{2}+2\,u+
2\,\pi -3), \nonumber\\
A_{1,2}^{(2)}(u)&=&- \pi  {u}^{2}\,(u+1)^{2}({u}^{2}+2\,u+2\,\pi -3)((\pi  -2) {u}^{2} +4\,u+\pi -2) \times\nonumber\\
&\times&(({\pi}^{2}-4\,\pi +6) \,{u}^{4}+(2\,{\pi }^{2}-6\,\pi +8)\,{u}^{3}
-\nonumber\\
&-&(2\,{\pi }^{2}-20\,\pi +28)\,{u}^{2}+(2\,{
\pi }^{2}-6\,\pi + 8) u +{\pi }^{2}-4\,\pi +6), \nonumber\\
A_{-1,1}^{(1)}(u)&=&(\pi -2)(u+1)^{2}((2\,\pi - 3) {u}^{2}+2\,u +1), \nonumber\\
A_{-1,1}^{(2)}(u)&=&-\pi \,(u+1)^{2}
((2\,\pi - 3){u}^{2}+2\,u  +1)
((\pi - 2) {u}^{2} +4\,u +\pi  -2) \times\nonumber\\
&\times&(({\pi 
}^{2}-4\pi +6\,) u^4+
(2\,{\pi }^{2}-6\,
\pi +8)\,{u}^{3}-
(2\,{\pi }^{2}-20\,\pi +28)\,{u}^{2}+\nonumber\\
&+&(2\,{\pi }^{2}-6\,\pi +8)\,u+{\pi }^{2}-4\,\pi +6), \nonumber\\
A_{-2,2}^{(1)}(u)&=&- u (u^2 +(2 \pi - 2) u + 1)^{-1} (u+1)^{2} \times\nonumber\\
&\times&((\pi -6){u}^{4}-8\,{u}^{3}+(4\,{\pi }^{3} - 16\,{
\pi }^{2}-2\,\pi +28)\,{u}^{2}-8\,u+
\pi -6), \nonumber\\
A_{-2,2}^{(2)}(u)&=&- \pi \,u (u+1)^{2}
(({\pi }^{2}-4\,\pi +6)\,{u}^{4}+\nonumber\\
&+&(2\,{\pi }^{2}-6\,\pi +8)\,{u}^{3}-(2\,{\pi }^{2}
-20\,\pi +28)\,{u}^{2}+(2\,{\pi }^{2}-6\,\pi +8)\,u+{\pi }^{2}-
4\,\pi +6)^{2},\nonumber\\
A_{2,2}^{(1)}(u)&=&u (u^2 +(2 \pi - 2) u + 1)^{-1} ((2\,{\pi 
}^{2} -9\,\pi +      14) \,{u}^{6} + \nonumber\\
&+& (4\,{\pi }^{3}-20\,{\pi }^{2} +46\,\pi-28) \,{u}^{5} 
+( 12\,{\pi }^{3} -66\,{\pi }^{2} + 169\,\pi  -142)\,{u}^{4} - \nonumber\\
&-&(16\,{\pi }^{3} - 168\,{\pi }^{2}  + 412\,\pi - 312)\,{u}^{3}   
+ (12\,{\pi }^{3}     -66\,{\pi }^{2}   +169\,\pi    -142) \,{u}^{2}   + \nonumber\\
&+&(4\,{\pi }^{3}  - 20\,{\pi }^{2}  + 46\, \pi  -28)\,u    
 +2\,{\pi }^{2} -9\,\pi + 14), \nonumber\\
A_{2,2}^{(2)}(u)&=&-\pi \,u (u+1)^{2}
(({\pi }^{2}-4\,\pi +6)\,{u}^{4}+
(2\,{\pi }^{2}-6\,\pi +8)\,{u}^{3}-\nonumber\\
&-&(2\,{\pi }^{2}-20\,\pi +28)\,{u}^{2}+
(2\,{\pi }^{2}-6 \pi +8)\,u+{\pi }^{2}-
4\,\pi +6)^{2}
\nonumber
\end{eqnarray}


\newpage

\begin{thebibliography}{99}








%
\bibitem{palmer} R.G.Palmer, in: {\it Heidelberg Colloquium on Glassy Dynamics}, eds.:
J.L. van Hemmen and I.Morgenstern, lecture Notes in Physics {\bf 275}, (Springer,
Berlin, 1987), p.275
%
\bibitem{nakazato} K.Nakazato and K.Kitahara, Prog. Theor. Phys. {\bf 64}, 2261 (1980)
%
\bibitem{lebowitz2} P.Ferrari, S.Goldstein and J.L.Lebowitz, Diffusion, Mobility
and the Einstein Relation, in: {\it Statistical Physics and Dynamical Systems},
eds.: J.Fritz, A.Jaffe and D.Szasz (Birkh\"{a}user, Boston, 1985)
p.405
%
\bibitem{kehr} K.W.Kehr and K.Binder, in:
 {\it Application of the Monte Carlo Method in
Statistical Physics}, ed. K.Binder, 
(Springer-Verlag, Berlin, 1987) and references therein.
%
\bibitem{hughes}B.D.Hughes, {\it Random Walks and Random
Environments}, (Oxford Science Publ., Oxford, 1995)
%
\bibitem{bur2} 
S.F.Burlatsky,  G.Oshanin, M.Moreau and W.P.Reinhardt, Phys. Rev. E {\bf 54},  3165 (1996) 
%
\bibitem{benichou} O.B{\'e}nichou, A.M.Cazabat, A.Lemarchand,
 M.Moreau and G.Oshanin, J. Stat. Phys. {\bf 97}, 351 (1999)
%
\bibitem{granek} R.Granek and A.Nitzan, J. Chem. Phys. {\bf 92}, 1329 (1990)
%
\bibitem{klafter} O.B{\'e}nichou, J.Klafter, M.Moreau and G.Oshanin,
Phys. Rev. E {\bf 62}, 3327 (2000)
%
\bibitem{saar1} R. van Gastel, E.Somfai, S.B. van Albada, W. van Saarloos and
J.W.M.Frenken, to appear in Phys. Rev. Lett; cond-mat/0009436
%
\bibitem{saar2} J.W.M. Frenken, R. van Gastel, S.B. van Albada, E. Somfai, W. van 
Saarloos, in: {\it Proc. NATO Advanced Research Workshop on "Collective
Diffusion on Surfaces: Collective Behaviour and the Role of
Adatom Interactions"}, Eds. M.C. Tringides and Z. Chvoj, (Kluwer, Dordrecht)
%
\bibitem{saar3} R. van Gastel, E. Somfai, W. van Saarloos and J. W. M. Frenken, 
Nature {\bf 408}, 665 (2000)
%
\bibitem{hilhorst1} M.J.A.M. Brummelhuis and H.J.Hilhorst, J. Stat. Phys. {\bf 53}, 249
(1988); see also: Marco Brummelhuis, {\it Application of Random Walk: Tracer
Diffusion, Lattice Covering, and Damage Spreading}, Ph. D. Thesis, Leiden University,
The Netherlands, 1991 
%
\bibitem{newman} T.J.Newman, Phys. Rev. B {\bf 59}, 13754 (1999)
%
\bibitem{toro} Z.Toroczkai, Int. J. Mod. Phys. {\bf B11}, 3343 (1997)
%
\bibitem{zia} R.K.P.Zia and Z.Toroczkai, J. Phys. A {\bf 31}, 9667 (1998)
%
\bibitem{hilhorst2} M.J.A.M. Brummelhuis and H.J.Hilhorst, Physica A {\bf 156}, 575
(1989)
%
\bibitem{ajay}  Ajay and R.G.Palmer, J. Phys. A {\bf 23}, 2139 (1990)
%
\bibitem{olla} C.Landim, S.Olla and S.B.Volchan, 
Commun. Math. Phys. {\bf 192}, 287 (1998)
%
\bibitem{bur3} G.Oshanin, J.De Coninck, M.Moreau and S.F.Burlatsky, 
 in: {\it Instabilities and
Non-Equilibrium Structures VII}, ed. E Tirapegui, 
(Kluwer Academic Pub., Dordrecht), to be published; 
cond-mat/9910243
%
\bibitem{feller} W.Feller, {\it An Introduction to Probability Theory and its Applications}, (Wiley, New York, 1970), Vols.1,2, 2nd
edition
%
\bibitem{vankampen} N.G.van Kampen, {\it Stochastic Processes in Physics and
Chemistry},  (North-Holland, New York, 1981)
%
\bibitem{montroll} E.W.Montroll, J. Math. Phys. {\bf 10}, 753 (1969); Proc. Symp. Appl. Math. {\bf 16}, 193 (1964)
%
\bibitem{mcrea} W.McCrea and F.Whipple, Proc. R. Soc. (Edinburg) 
{\bf 60}, 281 (1940) 
%
\bibitem{spitzer} F.Spitzer, {\it Principles of
 Random Walk}, (Van Nostrand,  Princeton, New Jersey, 1964)
%
\bibitem{benichou2}  O.B\'enichou, A.M.Cazabat, J.De Coninck, M.Moreau and G.Oshanin, Phys. Rev. Lett. {\bf 84}, 511 (2000)
%
\bibitem{cugl1} J.Kurchan, J. Phys. A {\bf 31}, 3719 (1998)
%
\bibitem{cugl2} L.F.Cugliandolo, D.S.Dean and J.Kurchan, Phys. Rev. Lett. {\bf 79}, 2168 (1997)  
%
\bibitem{sinai} Ya.B.Sinai, Theory Probab. Appl. {\bf 27}, 247 (1982); 
H.Kesten, M.V.Kozlov and F.Spitzer, Compositio Math. {\bf 30}, 145 (1975); 
B.Derrida and Y.Pomeau, Phys. Rev. Lett. {\bf 48}, 627 (1982) 


\end{thebibliography}
\end{document}